\documentstyle[aps,prl]{revtex}
\input epsf.sty
\begin{document}
\twocolumn[\hsize\textwidth\columnwidth\hsize\csname@twocolumnfalse\endcsname

\draft

\title{Excitation spectrum and effective mass of the even-fraction\\
quantum Hall liquid}
\author{Masaru Onoda}
\address{Institute of Particle and Nuclear Studies,\\
High Energy Accelerator Research Organization, Tanashi,
Tokyo 188-8501, Japan}
\author{Takahiro Mizusaki, Takaharu Otsuka, and Hideo Aoki}
\address{Department of Physics, University of Tokyo,
Hongo, Tokyo 113-0033, Japan}

\date{\today}

\maketitle

\begin{abstract}
To probe the nature of the even-fraction quantum Hall system,
we have investigated the low-lying excitation spectrum 
by means of exact diagonalization for finite systems.  We have found 
(i) a striking one-to-one correspondence (i.e., a shell 
structure) between the spectrum and those for free (composite) fermions, 
(ii) a surprisingly straight scaling plot for the excitation 
energy that gives a zero gap (metal) in the thermodynamic limit, 
(iii) the effective mass evaluated from the scaling 
becoming heavier for $\nu=1/2, 1/4, 1/6$, but 
(iv) some deviations from the single-mode  
or the Hartree-Fock composite fermion approximation as well.  

\end{abstract}


]

In the physics of the fractional quantum Hall system,
the composite fermion (CF) picture \cite{CF}
not only serves as an illuminating way of understanding 
Laughlin's incompressible 
quantum liquid for the odd-fraction Landau level filling, $\nu$, 
but also poses an interesting question of what is the
nature at even fractions, which is
the accumulation point of the fractional quantization.
A seminal paper by Halperin, Lee, and Read\cite{RPA}
suggested that the system at $\nu =1/2$ should be
a Fermi liquid of CF's in the mean-field picture, which 
led to intensive studies.  In contrast to the incompressible 
quantum Hall state or superconductors where energy gaps arise from many-body 
effects, we have to question here how the 
gap vanishes (i.e., how the liquid becomes compressible) 
despite the presence of the electron correlation.

Naively a CF, composed of
an electron and an even number ($\tilde{\phi}=2, 4, \cdots$) of flux quanta,
feels the mean magnetic field
$B_{\rm eff}=(\nu^{-1}-\tilde{\phi})\phi_0\rho$,
where $B=\nu^{-1}\phi_0\rho$ is the external magnetic field,
$\rho$ the number density of electrons,
and $\phi_0\equiv2\pi/e$ the flux quantum
(in the units in which $c=1$ and $\hbar=1$).
Thus $B_{\rm eff}$ vanishes for $\nu=1/\tilde{\phi}$.  
There is, however, no guarantee that the mean field should be 
good, and the above argument does not in fact say anything as to 
where the electron-electron repulsion comes in.  
Recent developments \cite{Read,Rajaraman_Sondhi,Shankar_Murthy} 
have indicated that we can define a `dipole' 
(composite particle + a correlation hole), where the flux-attachment is 
thought to mimic the repulsive correlation of electrons.  
The Halperin-Lee-Read prediction on $\nu =1/2$ has been
re-examined in the dipole picture, and the compressible nature is
reproduced \cite{SHOS}.

These approaches still adopt 
mean-field treatments, 
and their validity has yet to be fully clarified.  The 
difficulty arises because fluctuations of
the Chern-Simons gauge field that implements the flux-attachment
should be significant. The fluctuations in fact determine
the residual interaction between CF's as well as
the effective mass, $m^*$, of a CF, which are difficult to 
evaluate analytically.
Hence exact numerical studies for finite systems are valuable.  
Rezayi and Read \cite{Rezayi_Read} have numerically shown
that the ground state for the $\nu=1/2$ system on a sphere
has the same angular momentum as expected from
Hund's second rule for the same number of fermions in $B=0$.  
Morf and d'Ambrumenil \cite{Morf_d'Ambrumenil} have 
estimated $m^*$ from the size scaling of the ground-state energy.  
However, we are still some way from understanding
to what extent the CF picture applies.

One direct way going beyond the ground state is to look at 
the low-lying excitation spectrum --- 
here we question whether or not there is a {\it one-to-one correspondence}, 
in the structure of the excitation spectrum, 
between the $\nu=1/2$ liquid and a free
fermion system in $B=0$.  This can also enable us to extract, thorough 
the size-scaling of the energy gap, the effective mass.
This is exactly the motivation of the present work.

There are two points we wish to make:
(i) how to perform the size scaling is always a subtle problem, 
especially so when detecting the excitation gap
that may vanish in the thermodynamic limit.  
(ii) some analytic studies\cite{RPA,RG} have indicated 
that the nature of the $\nu=1/2$ liquid is affected by 
the range of the electron-electron interaction.
So we have taken a specific scaling sequence, and also
varied the range in monitoring the excitation spectrum.

We shall show that, (i) we do have a striking 
one-to-one correspondence between
the interacting and free systems. The shell structure in the 
spectrum is deformed with the range of the interaction, which is interpreted 
here in terms of the single-mode approximation (SMA), 
(ii) the effective mass becomes heavier as $\nu=1/2\rightarrow 
1/4\rightarrow 1/6$, somewhat more slowly than the Hartree-Fock (HF)
prediction of $m^*_{\rm HF} \propto 1/\nu^2$.  

We adopt the edge-free spherical system following Haldane,
which has an extra virtue
that the full rotational symmetry can be exploited in classifying the states.
Dirac's quantization condition dictates that
the total flux $4\pi R^2 B$ be an integral ($2S$) multiple of 
the flux quantum, where $R$ is the radius of the sphere. 
The eigenvalue of the non-interacting part of the Hamiltonian is
$\varepsilon = [l(l+1)-S^2]/(2mR^2)$, where $l$ ($\ge S$) is an integer.
The lowest Landau level (LLL) corresponds to $l=S$ with 
the Landau level filling given by $\nu=(N-1)/2S$ for $N$ electrons.

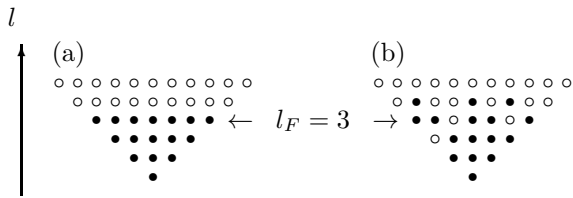
\begin{figure}[h]
\setlength{\unitlength}{0.5mm}
\begin{picture}(150,50)(0,0)
\put(72.5,22.5){\makebox(20,5)
{$\leftarrow\hspace{.2cm} l_F=3\hspace{.2cm}\rightarrow$}}
\put(0,50){\makebox(5,5){$l$}}
\put(5,5){\vector(0,1){40}}
\put(15,40){\makebox(5,5){(a)}}
\multiput(15,35)(5,0){11}{\circle{2}}
\multiput(20,30)(5,0){9}{\circle{2}}
\multiput(25,25)(5,0){7}{\circle*{2}}
\multiput(30,20)(5,0){5}{\circle*{2}}
\multiput(35,15)(5,0){3}{\circle*{2}}
\put(40,10){\circle*{2}}
\put(100,40){\makebox(5,5){(b)}}
\multiput(100,35)(5,0){11}{\circle{2}}
\put(105,30){\circle{2}}
\put(110,30){\circle*{2}}
\multiput(115,30)(5,0){2}{\circle{2}}
\put(125,30){\circle*{2}}
\put(130,30){\circle{2}}
\put(135,30){\circle*{2}}
\multiput(140,30)(5,0){2}{\circle{2}}
\multiput(110,25)(5,0){2}{\circle*{2}}
\put(120,25){\circle{2}}
\multiput(125,25)(5,0){2}{\circle*{2}}
\put(135,25){\circle{2}}
\put(140,25){\circle*{2}}
\put(115,20){\circle{2}}
\multiput(120,20)(5,0){4}{\circle*{2}}
\multiput(120,15)(5,0){3}{\circle*{2}}
\put(125,10){\circle*{2}}
\end{picture}
\caption{(a) A closed-shell ground state of the $N=16$ free system.
Solid (open) circles represent occupied (empty) states.
(b) An example of multi-exciton excitations
($[l_F-1][l_F]^2 \to [l_F+1]^3$ here).}
\label{free}
\end{figure}

For later reference let us look at the
spectrum for free fermions in zero magnetic field,
which is the mean-field solution for CF's at $\nu=1/2$.
The energy of a free fermion on a sphere is readily
given by $\varepsilon = l(l+1)/(2m^*R^2)$, where $l (\ge0$) is
the angular momentum and $m^*$ the fermion's mass.
We have to note that, with each level being $(2l+1)$-fold
degenerate, a `closed shell' configuration is realized when
$N = (l_F+1)^2$ [Fig.~\ref{free}(a)].
Here $l_F (=1, 2, 3, \cdots$ for $N=4, 9, 16, \cdots$)
is the highest occupied $l$,
so that we may call this the Fermi angular momentum
in analogy with the Fermi momentum in the planar system.

When $N\ne(l_F+1)^2$, the ground state of the non-interacting system
is thus degenerate, or has an `open shell'.
For interacting particles the degeneracy is lifted, and 
the total angular momentum
of the ground state becomes nonzero \cite{Rezayi_Read}.  
Since this can obscure the scaling, 
a more straightforward way is to
concentrate on the closed-shell sequence satisfying 
$N=(l_F+1)^2 = 4, 9, 16, \cdots$.
For this sequence the total angular momentum of the ground state
remains zero, and
provides a natural sequence toward the infinite system
for establishing both the structure of the low-lying
excitation spectrum and the energy gap.

The simplest class of excitations from a closed shell 
is `single-exciton' excitations where 
a particle is ejected from the $l_F$-th shell to 
the $(l_F+1)$-th, as has been pointed out by Rezayi and 
Read \cite{Rezayi_Read}.
The exciton's angular momentum takes the values $L= 1, \cdots, 2l_F+1$.
These excitations (abbreviated here as $[l_F]\to [l_F+1]$) 
provide the lowest-lying branch for $1\le L \le 2l_F+1$.

\begin{figure}[h]
  \begin{picture}(200,150)
    \put(0,0){\epsfxsize 200pt \epsfbox{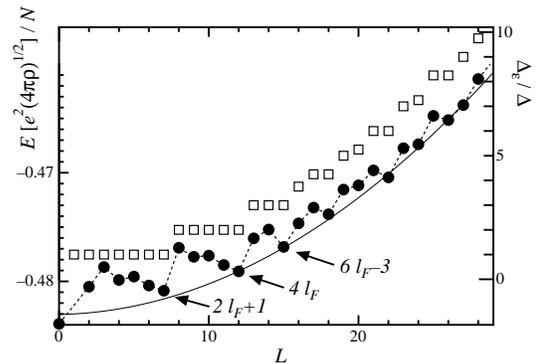}}
  \end{picture}
  \caption{Low-lying excitation spectrum for $N=16$ (solid circles).  
  Dashed line is a guide for the eye, while the full curve 
  represents $L(L+1)$. The low-lying excitation spectrum 
  $\Delta_{\varepsilon}$ for free fermions is also shown ($\Box$). 
  }
\label{16e}
\end{figure}

We can generalize this, including {\it multiple excitons}, 
to obtain the whole picture.  
For $2l_F+1 < L \le 4l_F$, the lowest-lying excitations are 
$[l_F]^2 \to [l_F+1]^2$, i.e., two-exciton excitations. 
For $4l_F < L \le 6l_F-3$ for $N\ge 9$ 
$[l_F]^3 \to [l_F+1]^3$ and so on, where $n$-exciton excitations 
$[l_F]^n \to [l_F+1]^n$ exist for $L \leq n$ $(2l_F+2-n)$.
For $L \ge 6l_F-2$, more complicated excitations such as
$[l_F-1]^m [l_F]^n \to [l_F+1]^{m'}[l_F+2]^{n'}$
($m+n=m'+n'$) must also be considered.  
Overall, however, 
the lowest-lying states are one-, two-, three-, $\cdots$ excitons, 
whose energies $\Delta_{\varepsilon}$ form {\it steps} moving up 
at $L_{\rm MAX} \simeq 2l_F, 4l_F, 6l_F, \cdots$, 
respectively, 
as shown in Fig.~\ref{16e} for $N=16$, 
although there are finite-size corrections in 
$L_{\rm MAX} = 2l_F+1, 4l_F, 6l_F-3$, etc as we have seen.  

Having looked at the free case, we now come to 
the structure of low-lying excitations 
in the interacting system.
The exact low-lying energies are 
obtained by diagonalizing the Hamiltonian matrix.
For $\nu=1/2$ we have $2S=2(N-1)$, 
and the dimension of the Hamiltonian is 4,669,367 
in the $L_z=0$ subspace for $N=16$ electrons.
The matrix elements can be expressed
in terms of Haldane's LLL projected pseudo-potential,
$V_J$ \cite{Haldane,FOC}.
If we explore the evolution of the spectra to $N=16$ (Fig.~\ref{16e}) 
from those for $N=4$ (not shown) and 
$N=9$ (Fig.~\ref{spectra} below), we are led to a well-defined 
series of cusps in the excitation spectrum, 
whose positions {\it exactly} agree with the above-mentioned positions
for the free fermions.  The degeneracies in the latter case 
are naturally lifted due to the interaction between CF's, but 
the interaction is weak enough to preserve the 
shell structure, which remarkably persists up to 
the angular momentum as large as 30 \cite{commentL1}. 
This is the first key finding in this Letter.

\begin{figure}[h]
  \begin{picture}(200,200)
    \put(0,0){\epsfxsize 200pt \epsfbox{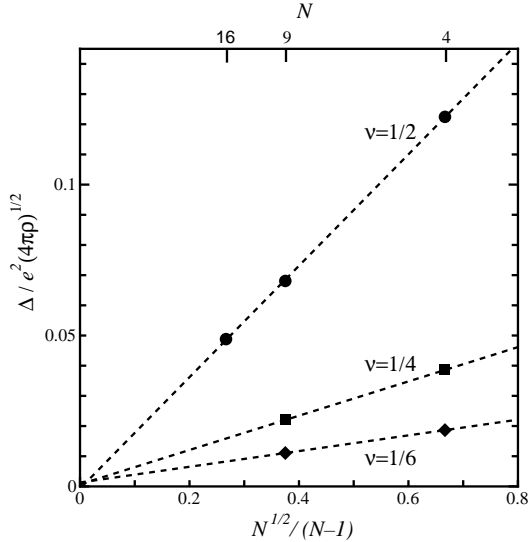}}
  \end{picture}
  \caption{Size scaling of the gap for $\nu=1/2, 1/4, 1/6$. 
  The dashed lines are linear fit to the data.  }
\label{gap}
\end{figure}

We can next evaluate the energy gap, $\Delta$.  
In the free fermion system, the lowest excitation corresponds to 
$l_F \to l_F+1$ with 
$\Delta\equiv (l_F+1)/(m^*R^2)$. This quantity has a well-defined
scaling, $\Delta = (4\pi\rho/m^*)[\sqrt{N}/(N-1)]$
when $N$ is varied with $\rho$ fixed.
For the interacting system, the cusped structure 
revealed here enables us to identify the position of 
the lowest excited state, which always occurs for the first cusp
at $L=2l_F+1$ (the high-$L$ end of the single-exciton excitation).
Fig.~\ref{gap} shows this gap for $\nu=1/2$ \cite{N4L3}.
We can immediately see a surprisingly accurate
linear scaling that extrapolates to zero
for $N\rightarrow \infty$ if we have $\sqrt{N}/(N-1)$ as the
scaling variable, as guided by the free-system behavior, 
$\Delta = (4\pi\rho/m^*)[\sqrt{N}/(N-1)]$

This same formula can be used to extract
the effective mass $m^*$ of CF's from the gradient of the scaling
plot, with the result 
$1/m^* = (0.185 \pm 0.002) e^2\ell$,
where ${\ell}\equiv 1/\sqrt{eB}$ is the magnetic length.
The $1/m^*$ obtained here from the excitation gap is
slightly smaller than $1/m^*\simeq  (0.2\pm0.02)e^2\ell$,
estimated from the ground-state energy per particle\cite{Morf_d'Ambrumenil}.
On the other hand the present value is slightly larger than the 
analytic estimate, $1/m^*\simeq e^2\ell/6$, obtained from the 
interaction energy between an electron and a correlation-hole
in the first-quantized picture \cite{Shankar_Murthy}
or the self-energy of the CF in the temporal gauge
in the HF approximation \cite{YSD}.

\begin{figure}[h]
  \begin{picture}(200,140)
    \put(0,0){\epsfxsize 200pt \epsfbox{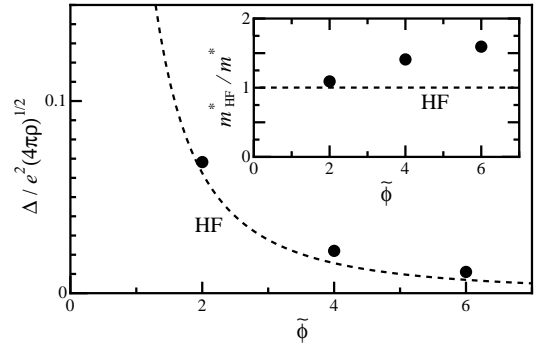}}
  \end{picture}
  \caption{$\Delta$ ($\propto 1/m^*$) for the sequence 
  $\nu = 1/2, 1/4, 1/6$ for $N=9$. The dashes lines represent 
  the HF result, with the inset depicting $m^*_{\rm HF}/m^*$.  
}
\label{even}
\end{figure}

The gap and mass, dominated by gauge field fluctuations, should 
depend on the number of flux quanta attached ($\tilde{\phi}$), so 
we further obtained the scaling plot for the sequence
$\nu = 1/\tilde{\phi} = 1/2, 1/4, 1/6$ in Fig.~\ref{gap} \cite{N9}.  
The gap again vanishes for $N\rightarrow \infty$, 
where the effective mass systematically becomes heavier 
in the sequence 
$\nu =1/2, 1/4, 1/6$ as shown in Fig.~\ref{even} for $N=9$.  
In the HF approximation for the CF we can show \cite{HF_mass} 
that $m^*$ should scale as
\begin{equation}
\frac{1}{m^*_{\rm HF}(\tilde{\phi})} =
\frac{1}{6}\left( \frac{2}{\tilde{\phi}} \right)^2
\frac{e^2}{\sqrt{4\pi \rho}}.
\end{equation}
This is a decreasing function of $\tilde{\phi}$ as well, 
but the present numerical result is seen to deviate from the HF 
result (inset of Fig.~\ref{even}) for larger $\tilde{\phi}$.\cite{exp14}

Now let us look more closely at the excitation spectra.  
Note in passing that the overall shape of the spectrum
exhibits an $\propto L^2$ asymptote 
as evident from Fig.~\ref{16e}.  We can explain this by 
converting the Hamiltonian in the $c^{\dagger}c^{\dagger}cc$ form
to $c^{\dagger}cc^{\dagger}c$. We have then, up to a constant, 
$
e^2/(2{\ell} \sqrt{S})
\sum^{2S}_{K=1}\tilde{V}_K \:\rho_{K}\cdot\rho_{K}
$
where 
$\rho_{K}\cdot\rho_{K}\equiv \sum (-1)^M \rho_{KM}\:\rho_{K,-M}$
and 
$\rho_{KM}\equiv \sum
(-)^{S+m_2}\langle Sm_1; Sm_2 | KM \rangle \:
c_{m_1}^{\dagger}c_{-m_2}$ where $c_{m}^{\dagger}$ creates the $m$th orbit.  
The transformed coefficient becomes
$\tilde{V}_K \equiv
\sum_{J=0}^{2S} (-1)^{2S+J}(2J+1)
\{ ^{SSJ}_{SSK}\} V_J$,
where $\langle j_1 m_1; j_2 m_2 | JM \rangle$
is the Clebsch-Gordan coefficient and $\{ ^{SSJ}_{SSK}\}$ 
Wigner's $6j$ symbol. In this representation, 
while $\rho_{1M}$ is nothing but
the (LLL projected) total angular momentum operator, 
so the leading term becomes
$\rho_1\cdot\rho_1 = [3/S(S+1)(2S+1)]
\hat{\mbox{\boldmath $L$}}\cdot\hat{\mbox{\boldmath $L$}}$,
which explains the asymptote $\propto L(L+1)$.

Now we come to what happens when the range of interaction is changed.
We have calculated the excitation spectra
replacing the pseudo-potential $V_{2S-m}$ with $(V_{2S-m})^a$.
Since $V_{2S-m}$ is the potential between two electrons with the
relative angular momentum $m$, 
$a<1$ $(a>1)$ corresponds to the interaction longer- (shorter-)ranged
than Coulombic.  

The numerical result in Fig.~\ref{spectra} \cite{commentL1} shows that 
the cusped structure in the spectra becomes more pronounced 
(i.e., effect of the inter-CF interaction becomes enhanced) 
as the interaction is made shorter-ranged, 
although the positions of cusps remain the same.  
So the free CF picture seems to be better for longer-ranged
interaction.  This is in sharp contrast with the Laughlin's liquid at
odd denominators for which the mean-field CF picture yields even
an exact ground state when the interaction is short-ranged enough.
The cusps sticking to $2l_F, 4l_F, \cdots$
remind us of the Tomonaga-Luttinger(TL) liquid, which is a totally different 
system in one dimension, but the cusps, having a 
topological origin, do not change with the form of interaction.

\begin{figure}[h]
  \begin{picture}(200,200)
     \put(0,0){\epsfxsize 200pt \epsfbox{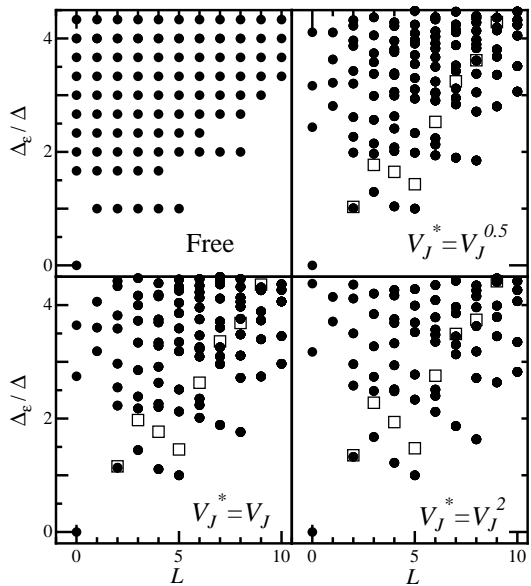}}
  \end{picture}
  \caption{
Full excitation spectra for $\nu=1/2$ with
$(V_{2S-m})^a$ ($a = 0.5, 1.0, 2.0$) for $N=9$ ($l_F=2$). The
energy is normalized by the gap at $L=2l_F+1 (=5)$ for
each value of $a$. The SMA result is also shown ($\Box$).
}
\label{spectra}
\end{figure}

The tendency that the system lies the further away from the Fermi liquid
the shorter-ranged the interaction 
is consistent with analytic studies.  Namely, 
an improved random-phase approximation (RPA) \cite{RPA}
and a renormalization group (RG) study \cite{RG}
suggest that for short-ranged potential 
the one-particle Green's function
has a branch cut rather than a pole just as in the TL liquid.  
For longer-ranged case the Fermi-liquid properties are recovered.
To test these predictions from numerical low-lying spectra
will require further investigations, including correlation function studies.
However, we can compare the behavior of the lowest 
cusp (i.e., single-exciton branch) with the SMA, where 
the $\rho_{LM}$ defined above operated 
on the ground state $|\Psi_0\rangle$ 
is used as the trial function in evaluating the energy, 
$=\langle\Psi_0|\rho_{LM}^{\dagger}(H-E_0)\rho_{LM}|\Psi_0\rangle/
\langle\Psi_0|\rho_{LM}^{\dagger}\rho_{LM}|\Psi_0\rangle = f(L)/s(L)$.  
The SMA result ($\Box$ in Fig.~\ref{spectra}) roughly reproduces the 
gradient of the branch, although we encounter a deviation 
larger than those in the odd-fraction liquids.  
We can numerically show that 
the structure factor $s(L)$ remains almost identical as the interaction 
range is varied, so the change in the oscillator strength 
$f(L)$ is dominating the shape of the cusp.  

To summarize, the present numerical result, done on the 
largest scale currently available, has enabled us to show that 
the gauge fluctuations in the even-fraction metals 
are substantial, but not so strong as to destroy the shell structure 
in the low-lying excitation spectrum.  
We are also extending the present study to the spin degrees of freedom, 
which will be published elsewhere.
We acknowledge Peter Maksym for a critical reading of the manuscript, and 
Kazuhiko Kuroki for illuminating discussions.



\begin{references}
\bibitem{commentL1} 
For $L=1$ single-exciton states cannot be constructed as shown by SMA
\cite{Rezayi_Read}, so we have excluded the $L=1$ result 
in Fig.~\ref{16e} but included in Fig.~\ref{spectra}.
\bibitem{N4L3}
For $N=4$ and $\nu=1/2$, we have plotted 
the excitation energy for $L=3 (=2l_F+1)$
to compare with the $N=9, 16$ results for which
the lowest excitations have $L=2l_F+1$.
\bibitem{N9} The dimension of the Hamiltonian for $\nu=1/6, N=9$ is 
about 20 million, which is as far as we can treat with a 
parallel processing.
\bibitem{HF_mass}
Ref. \cite{YSD} also reports the HF mass,
but the factor involving $\tilde{\phi}$ is not explicitly included.
\end{references}
\end{document}